\documentclass{article}
\usepackage{cite}
\usepackage{graphicx}

\begin{document}

\newcommand{\bear}{\begin{eqnarray}}
\newcommand{\eear}{\end{eqnarray}}
\newcommand{\be}{\begin{equation}}
\newcommand{\ee}{\end{equation}}
\newcommand{\beqn}{\begin{eqnarray}}
\newcommand{\eeqn}{\end{eqnarray}}
\newcommand{\beqnn}{\begin{eqnarray*}}
\newcommand{\eeqnn}{\end{eqnarray*}}

\newcommand{\exponente}[1]{\mathrm{e}^{#1}}
\newcommand{\adj}[1]{\left( #1 \right)}
\newcommand{\X}{\hat{x}}
\newcommand{\p}{\hat{p}}

\newcommand{\avg}[1]{\langle#1\rangle}

\def\vep{\varepsilon}
\def\vf{\varphi}
\def\al{\alpha}

\begin{center} {\Large \bf
Adiabatic amplification of the harmonic oscillator energy 
\\
when the frequency passes through zero}

\end{center}

\begin{center} {\bf
Viktor V. Dodonov
and Alexandre V. Dodonov }
\end{center}
 
\begin{center}

{\it 
Institute of Physics,,
University of Brasilia, 
70910-900 Brasilia, Federal District, Brazil
\\
International Center for Physics, University of Brasilia, Brasilia, DF, Brazil }

\end{center}

{Correspondence: vdodonov@unb.br}

\abstract{
We study the evolution of the energy of a harmonic oscillator when its frequency slowly
varies with time and passes through zero value. We consider both the classical and quantum
descriptions of the system. We show that after a single frequency passage through zero value,
the famous adiabatic invariant ratio of energy to frequency (which does not hold for zero
frequency) is reestablished again, but with the proportionality coefficient dependent on the
initial state. The dependence on the initial state disappears  after averaging over phases
of initial states with the same energy (in particular, for the initial vacuum, Fock and thermal
quantum states). In this case, the mean proportionality coefficient
is always greater than unity. The concrete value of the mean proportionality coefficient 
depends on the power index of the frequency dependence
on time near zero point. In particular, the mean energy triplicates if the frequency tends
to zero linearly. If the frequency attains zero more than once, the adiabatic
proportionality coefficient strongly depends on lengths of time intervals between
zero points, so that the mean energy behavior turns out quasi-stochastic after many passages
through zero value. The original Born-Fock theorem does not work after the frequency passes 
through zero. However, its generalization is found: the initial Fock state becomes a wide
superposition of many Fock states, whose weights do not depend on time in the new adiabatic regime.
When the mean energy triplicates, the initial Nth Fock state becomes a superposition of,
roughly speaking, 6N states, distributed non-uniformly.
The initial vacuum and low-order Fock states become squeezed, as well as initial thermal
states with low values of the mean energy.
}

\section{Introduction}
\label{sec-intr}

One of many brilliant results of classical and quantum mechanics is the existence of 
{\em adiabatic invariants\/} in the case when parameters of a system vary slowly with time.
The simplest invariant is the ratio of the energy of a harmonic oscillator ${\cal E}(t)$ to its 
time-dependent frequency $\omega(t)$ \cite{Land-mech}:
\be 
{\cal E}(t)/ \omega(t) = const \quad \mbox{if} \; |\dot\omega|/\omega^2(t) \ll 1.
\label{inv}
\ee
Then, suppose that the frequency returns to its initial value after some slow variations. What will be
the final energy of the oscillator? According to Equation (\ref{inv}), the answer is obvious: the final energy 
will coincide with the initial one. However, there exists a remarkable exclusion from this result, when
the frequency passes through zero value in the process of evolution, so that the condition of validity of 
Equation (\ref{inv}) is obviously broken for any rate of the evolution. The goal of our paper is to study
the dependence of the final energy on the shape of time-dependent frequency, when this frequency 
slowly passes through zero value. 
We perform analytic calculations for the quantum oscillator
and numeric calculations for the classical oscillator.
Note that various aspects of the harmonic oscillator evolution in the adiabatic regime were studied by
many authors during decades (see, e.g., papers \cite{Kul57,Knorr66,Solim69,MMT73,Keller91,Robnik06,Robnik06e}). 
However, the situation
when the frequency passes slowly through zero value was not considered in the known publications.
An oscillator whose frequency exponentially goes adiabatically and asymptotically to zero was considered, e.g., in papers
\cite{Zaugg94,Moll97}. However, we are interested in the case when the frequency passes through zero and returns
to its initial value.

The structure of the paper is as follows.
In Section \ref{sec-clas}, we bring  results of numeric solutions of the classical equations
of motion for two time dependences of the frequency: $\omega^2(t) \sim |t|^n$ and 
$\omega^2(t) \sim |\tanh(at)|^n$.
Several figures show differences between the cases of fast and slow frequency variations,
when the frequency does not attain zero value and when it passes through zero value.
In Section \ref{sec-quantgen}, we bring general formulas describing the
evolution of the mean oscillator energy in the quantum case, including adiabatic
regimes without and with crossing zero frequency value.
Exact solutions for the power profile of the frequency are derived and
analyzed in Section \ref{sec-power}.
Transition rules in the case of single frequency passage through zero are obtained
in Section \ref{sec-trans}. An example of $\tanh$-like frequency functions is considered
in Section \ref{sec-EE}.
Double transitions of frequency through zero values are studied in Section \ref{sec-double}
in the quantum and classical cases.
Sections \ref{sec-fluct}, \ref{sec-Fock} and \ref{sec-sqz} are devoted to the energy fluctuations,
the violation and generalization of the Born--Fock theorem, and the appearance of squeezing,
respectively. Section \ref{sec-concl} contains the discussion of main results.

\section{Evolution of the oscillator energy in the classical case}
\label{sec-clas}

The basic equation has the form
\be
\ddot{x} + \omega^2(t) x =0.
\label{eq}
\ee
We consider a special case when the time-dependent frequency can be written as
$\omega^2(t) =\omega_0^2f(t/\tau)$, where $f(t/\tau)$ is a non-negative function with
the properties
\be
f(-1)=1, \qquad f(0)=0, \qquad \tau>0.
\label{propf}
\ee
We assume that $\omega^2(t)=\omega_0^2 = const$ for $t\le -\tau$. 
Since we are interested mainly in the evolution of energy at $t>-\tau$, 
we consider a one-parameter family of classical trajectories
with the same initial energy $E_0$. Then, assuming the particle mass $m=1$, 
it is convenient to use the initial coordinate
$x_0$ and initial velocity $\dot{x}_0$ in the form 
\[
x_0 = \cos(\vf) \sqrt{2E_0}/\omega_0, \qquad \dot{x}_0 = \sin(\vf) \sqrt{2E_0} , \qquad 0 \le \vf < 2\pi.
\]
To solve Equation (\ref{eq}) numerically, we introduce dimensionless variables $X= \omega_0 x/\sqrt{2E_0}$ and $T=t/\tau$,
arriving at the equation for $T\ge -1$
\be
d^2 X/dT^2 + G^2 f(T) X =0, \qquad G = \omega_0\tau,
\label{eq-XT}
\ee
with the initial conditions
\be
X(-1) = \cos(\vf), \qquad dX/dT|_{T=-1} = G \sin(\vf).
\label{inivf}
\ee
Then, the dimensionless energy ratio $R=E(t)/E_0$ depends on the dimensionless time $T$ and two parameters,
$G$ and $ \vf$:
\be
R(T;G; \vf) = f(T) X^2(T) + G^{-2}\left(dX/dT\right)^2 ,
\label{Rf}
\ee
where the derivative $dX/dT$ must be taken at instant $T$.
The existence of the adiabatic invariant (\ref{inv}) implies that 
$R(T;G)= \sqrt{f(T)}$ for $T<0$ and big enough values of parameter $G$,
{\em independently on the values of parameter $\vf$}.
In the following subsections we study, what can happen if $T>0$, for different families of
functions $f(T)$.

\begin{figure}[hbt]
\centering
\includegraphics[scale=0.4]{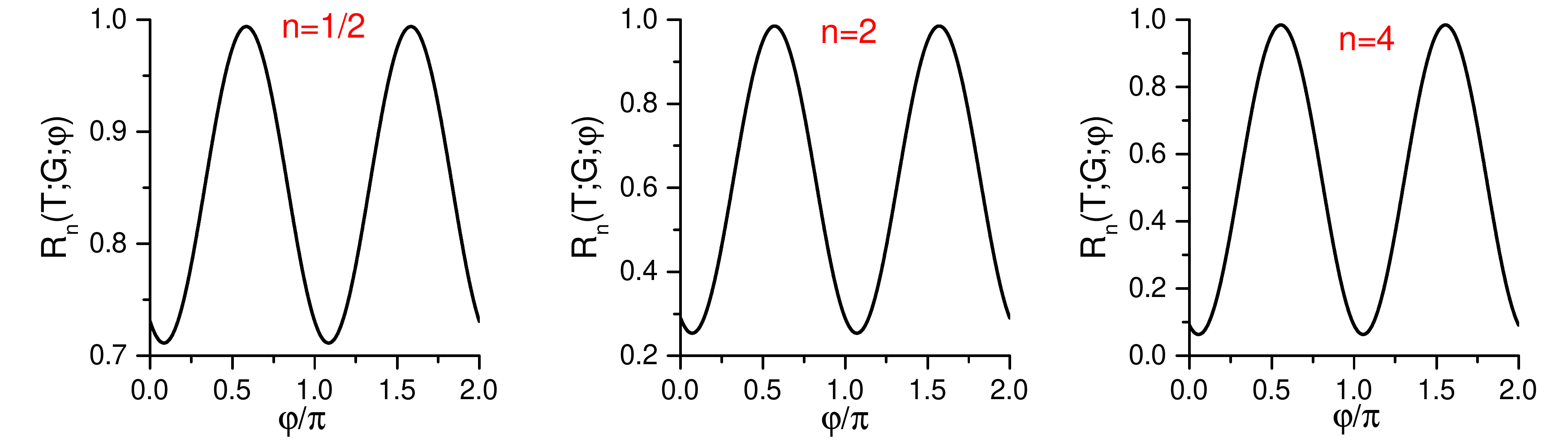}
%\vspace{0.5cm}
\caption{The dimensionless energy of a classical particle at the dimensionless instant $T=-1/2$ in the non-adiabatic regime ($G=1$).
%Left: for $n=1/2$. Middle: for $n=2$. Right: for $n=4$. 
}
\label{fig-T-half-G1}
\end{figure} 
\begin{figure}[hbt]
\centering
\includegraphics[scale=0.4]{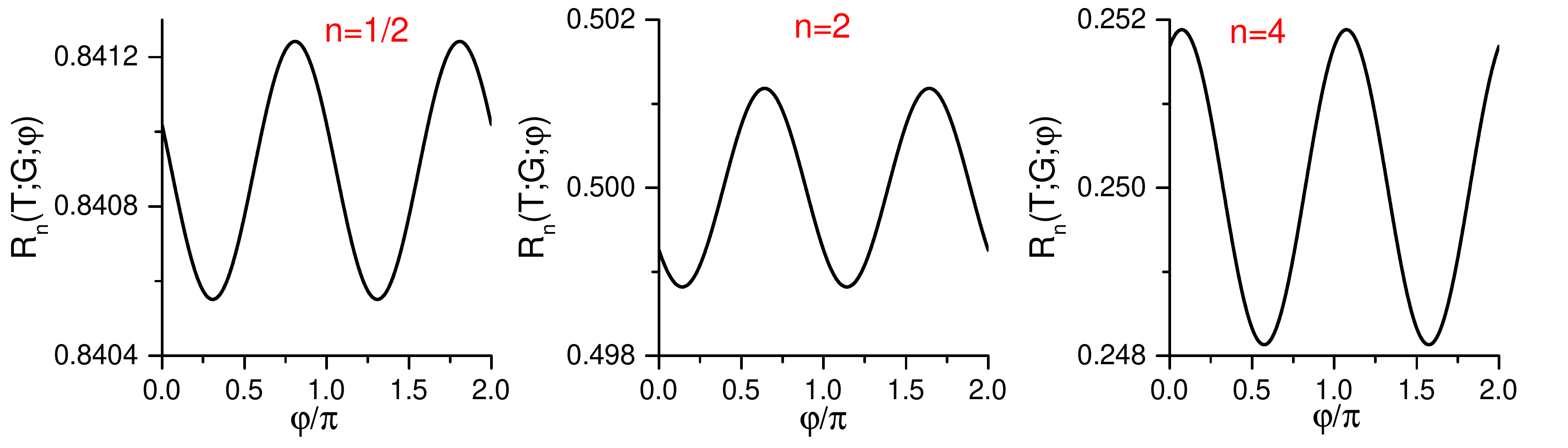}
%\vspace{0.5cm}
\caption{The dimensionless energy of a classical particle at the dimensionless instant $T=-1/2$ in the adiabatic regime ($G=1000$).
%Left: for $n=1/2$. Middle: for $n=2$. Right: for $n=4$. 
}
\label{fig-T-half-G1000}
\end{figure} 
\begin{figure}[hbt]
\centering
\includegraphics[scale=0.4]{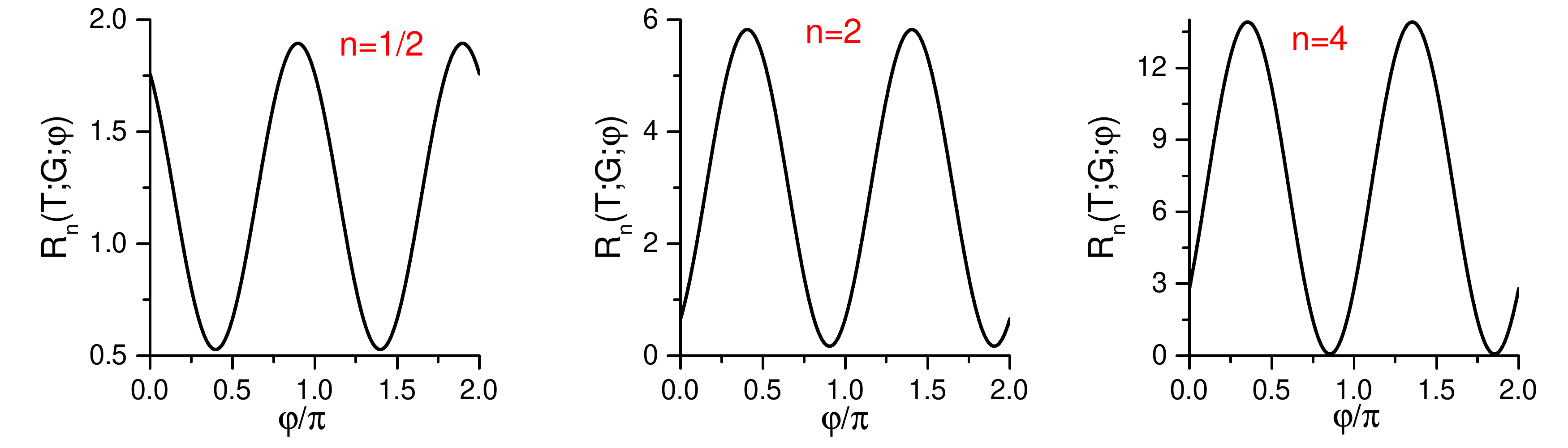}
%\vspace{0.5cm}
\caption{The dimensionless energy of a classical particle at the dimensionless instant $T=1$ in the case of slow evolution ($G=1000$), for the profile $f(T) = |T|^n$.
%Left: for $n=1/2$. Middle: for $n=2$. Right: for $n=4$. 
}
\label{fig-T1-G1000}
\end{figure} 
\subsection{A power profile of the frequency}

Our first example is the power profile $f(T) = |T|^n$ with $n>0$. 
We solved numerically Equation (\ref{eq-XT}) and calculated the dimensionless ratio
\[
R_n(T;G; \vf) = |T|^n X^2(T) + G^{-2}\left(dX/dT\right)^2.
\]
The adiabatic ratio must equal $R_n(T;G)= |T|^{n/2}$. 
In Figures \ref{fig-T-half-G1} and \ref{fig-T-half-G1000} we show
  $R_n$ as function of $\vf$ for the fixed values $T= -1/2$, $n=1/2$, $n=2$ and $n=4$, 
	comparing different behaviors when $G=1$ (no adiabaticity) and $G=1000$.
Figure \ref{fig-T-half-G1} shows a strong dependence of the energy on the phase $\vf$ 
in the non-adiabatic regime.
However, this dependence becomes negligible in Figure \ref{fig-T-half-G1000}, 
which shows that the mean value of $R_n$ is, indeed, $|T|^{n/2}$.
 But the situation becomes  quite different for $T\ge 0$: the dependence of $R_n$ on $\vf$ does not disappear even for very big values
 of parameter $G$. This is shown in Figure \ref{fig-T1-G1000}  for $R_n(1;1000;\vf)$
with $n=1/2$, $n=2$ and $n=4$. It is important to pay attention to different vertical scales in different plots.
Strong oscillations are observed. The mean value of these oscillations depends on index $n$. 

\subsection{A tanh-like profile of the frequency}

\begin{figure}[hbt]
\centering
\includegraphics[scale=0.4]{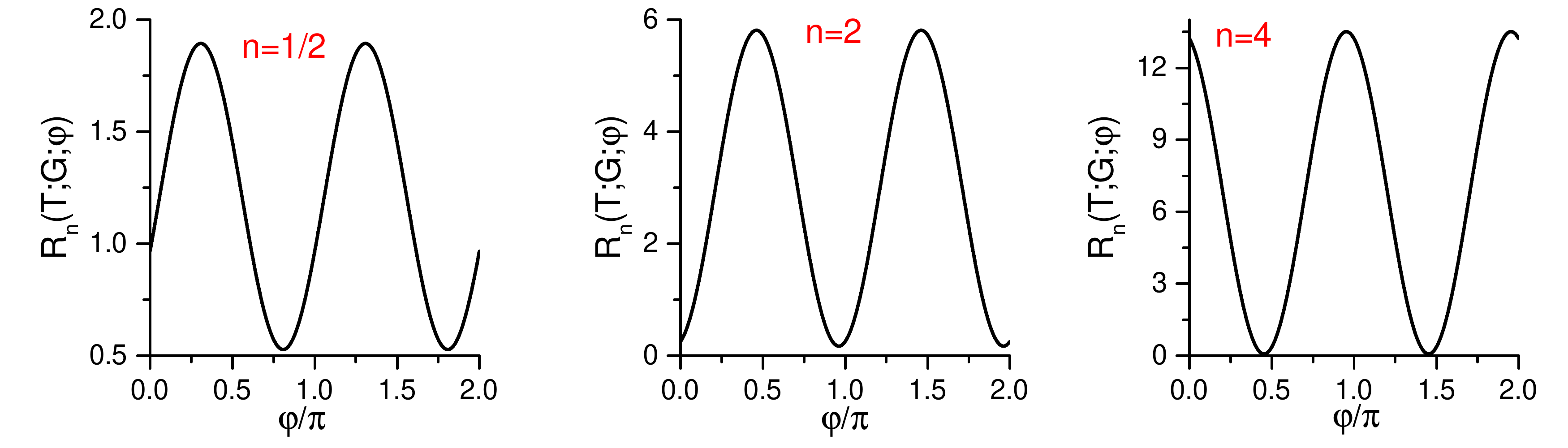}
%\vspace{0.5cm}
\caption{The dimensionless energy of a classical particle at the dimensionless instant $T=1$ in the case of slow evolution ($G=1000$) for the  profile $f(T) = |\tanh(aT)/\tanh(-a)|^n$ $a=5$.
}
\label{fig-T1-G1000-th}
\end{figure}

The second example is the  profile $f(T) = |\tanh(aT)/\tanh(-a)|^n$ with $n>0$ and $a=5$,
which describes a more soft transition from the constant frequency to a time-dependent one.
The plots in this case turn out very similar to those of the preceding section. We show one of them
in Figure \ref{fig-T1-G1000-th}, for
 the value $T=1$ (when the final frequency coincides with the initial one).
\begin{figure}[hbt]
\centering
\includegraphics[scale=0.25]{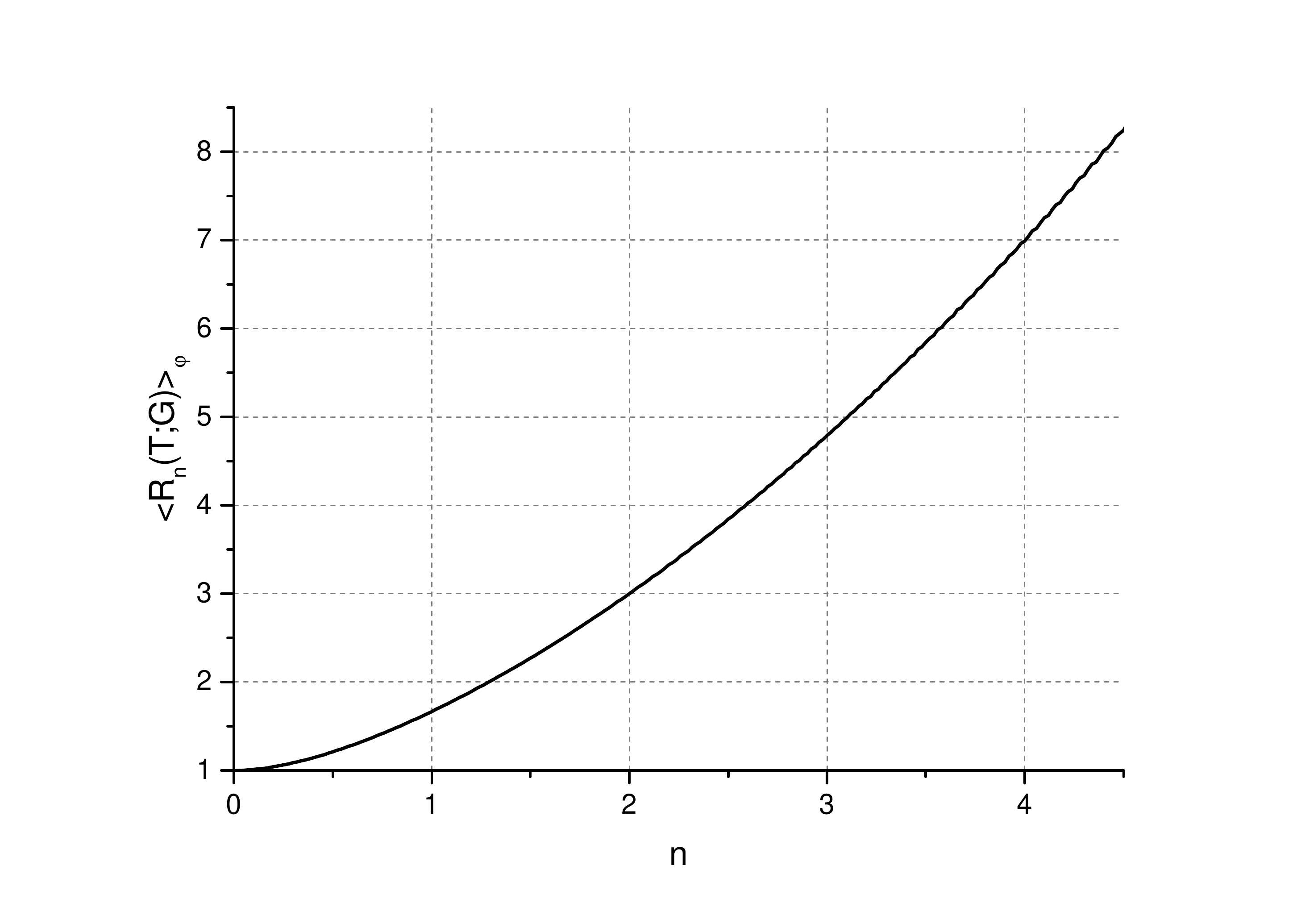}
\hfill
\includegraphics[scale=0.25]{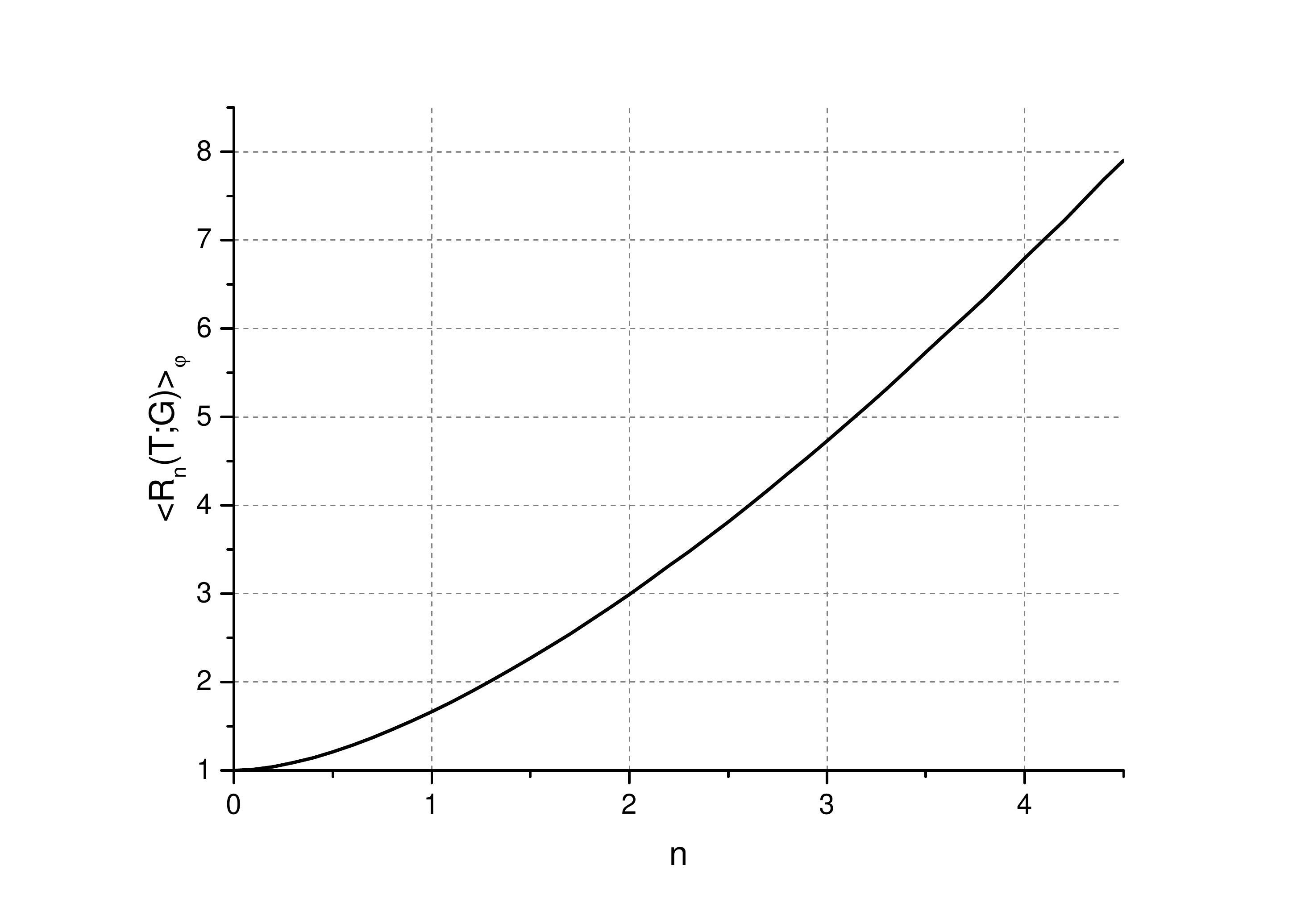}
\caption{The dimensionless energy of a classical particle at the dimensionless instant $T=1$ 
in the case of slow evolution ($G=1000$),
averaged over the initial phase $\vf$, as function of index $n$.
Left: for the profile $f(T) = |T|^n$. Right: 
for the  profile $f(T) = |\tanh(aT)/\tanh(-a)|^n$ $a=5$.
}
\label{fig-meanE}
\end{figure} 
While the lines $R(\vf)$ are shifted with respect to each other in the cases of $f(T) = |T|^n$
and $f(T) = |\tanh(aT)/\tanh(-a)|^n$, the maximal and minimal values coincide. Moreover, the average
values of the final energy as functions of index $n$ turn out practically identical
for two families of frequency profiles, as shown in Figure \ref{fig-meanE}.
This coincidence is explained in the following sections.

\section{Evolution of the mean oscillator energy in the quantum case}
\label{sec-quantgen}

In the quantum case, one has to solve the time-dependent Schr\"odinger equation and use the solutions to calculate
various mean values, in particular, those contributing to the mean energy. 
However, a more simple way is to use the Ehrenfest equations for the mean values,
which are immediate consequences of the Schr\"odinger equation. It was shown in 
 the seminal papers by Husimi \cite{Husimi}, Popov and Perelomov \cite{PP},
 Lewis and Riesenfeld \cite{LR}, and Malkin, Man'ko and Trifonov \cite{MMT70}, 
 that the solutions of the Schr\"odinger and Ehrenfest equations for the harmonic oscillator with an arbitrary
 time-dependent frequency depend on 
 complex functions  $\vep(t)$ and $\vep^*(t)$, satisfying Equation (\ref{eq}) and the initial conditions 
\be
\vep(-\tau)= \omega_{0}^{-1/2}, \quad \dot\vep(-\tau)= i\omega_{0}^{1/2}.
\label{invep}
\ee 
The Wronskian identity for the solutions $\vep(t)$ and $\vep^*(t)$ has the form
\be
\dot\vep \vep^* - \dot\vep^* \vep =2i.
\label{Wr}
\ee
Then, we can write at $t\ge \tau$
\be
x(t) = x_0 \sqrt{\omega_0}\, \mbox{Re}[\vep(t)] + \frac{p_0}{\sqrt{\omega_0}} \mbox{Im}[\vep(t)], \qquad
p(t) = x_0 \sqrt{\omega_0}\,\mbox{Re}[\dot\vep(t)] + \frac{p_0}{\sqrt{\omega_0}} \mbox{Im}[\dot\vep(t)].
\label{solxp}
\ee

Equation (\ref{solxp}) holds both for the classical and quantum particles (in the Heisenberg representation
in the latter case).
Its immediate consequences are  the following formulas for the 
second-order moments of the canonical operators for $t \ge -\tau$:
\be
\langle x^2\rangle_t =  \langle x^2\rangle_{-\tau}\, \omega_0 \left(\mbox{Re}[\vep(t)]\right)^2 + 
\frac{\langle p^2\rangle_{-\tau}}{\omega_0} \left(\mbox{Im}[\vep(t)]\right)^2
+ \langle xp + px \rangle_{-\tau} \left(\mbox{Re}[\vep(t)]\right) \left(\mbox{Im}[\vep(t)]\right),
\label{x2t}
\ee
\be
\langle p^2\rangle_t =  \langle x^2\rangle_{-\tau}\, \omega_0 \left(\mbox{Re}[\dot\vep(t)]\right)^2 + 
\frac{\langle p^2\rangle_{-\tau}}{\omega_0} \left(\mbox{Im}[\dot\vep(t)]\right)^2
+ \langle xp + px \rangle_{-\tau} \left(\mbox{Re}[\dot\vep(t)]\right) \left(\mbox{Im}[\dot\vep(t)]\right).
\label{p2t}
\ee
The time-dependent mean energy is given by the formula
\be
 {\cal E}(t) = \frac12\left[\langle p^2 \rangle_t 
+ \omega^2(t) \langle x^2 \rangle_t \right].
\label{defE}
\ee
It is worth remembering that for systems with {\em quadratic\/} Hamiltonians with respect to $x$
and $p$, the dynamics of the first-order mean values $\langle x\rangle$ and $\langle p\rangle$
are {\em totally independent\/} from the dynamics of the variances 
$\sigma_x = \langle x^2\rangle - \langle x\rangle^2$, 
$\sigma_p = \langle p^2\rangle - \langle p\rangle^2$ and
$\sigma_{xp} = \langle xp +px\rangle/2 - \langle x\rangle \langle p\rangle$.
This means that the equations of the same form as (\ref{x2t}) and (\ref{p2t}) exist for the
sets $(\langle x\rangle^2, \langle p\rangle^2, \langle x\rangle \langle p\rangle)$ and
$(\sigma_x, \sigma_p, \sigma_{xp})$.

The adiabatic (quasiclassical) approximate complex solution to Equation (\ref{eq}),
satisfying the initial conditions (\ref{invep}),
has the form
\be
\vep(t) \approx [\omega(t)]^{-1/2} e^{i\phi_{\tau}(t)}, 
\qquad \dot\vep(t) \approx i[\omega(t)]^{1/2} e^{i\phi_{\tau}(t)}, 
\qquad \phi_{\tau}(t) = \int_{-\tau}^t\omega(z)dz.
\label{adsol}
\ee
Putting the solution (\ref{adsol}) in the equations (\ref{x2t})-(\ref{defE}).
 we arrive immediately at the adiabatic invariant 
\be
{\cal E}(t)/\omega(t) = {\cal E}(-\tau)/\omega_0 , %\qquad -\tau < t <0.
\label{adiinv}
\ee
{\em for arbitrary initial values\/} at $t=-\tau$.
However, the solution (\ref{adsol}) obviously looses its sense if $\omega(t) =0$ 
at some time instant $t_*$ (taken as $t=0$ in our paper).
Nonetheless, when the frequency slowly passes through zero value and slowly becomes not too small, 
the conditions of the quasiclassical approximation are reestablished again. 
Hence, the solution for $t>0$  can be written (outside some interval near $t=0$)
in the most general quasiclassical form as follows,
 \be
\vep(t) \approx [\omega(t)]^{-1/2} \left[ u_{+} e^{i{\phi}(t)} + 
 u_{-} e^{-i{\phi}(t)} \right], 
\quad \dot\vep(t) \approx i[\omega(t)]^{1/2} 
\left[ u_{+} e^{i{\phi}(t)} -  u_{-} e^{-i{\phi}(t)} \right],
\label{adsol+}
\ee
where
\be
\phi(t) = \int_{t_*}^t\omega(z)dz, \qquad d\phi(t)/dt =\omega(t).
\label{t*}
\ee
Constant complex coefficients $u_{\pm}$ must obey the condition  
\be
|u_{+}|^2 - |u_{-}|^2 =1,
\label{uvcond}
\ee
which is the consequence of Equation (\ref{Wr}). 
Then, Equation (\ref{defE}) assumes the form
\be
\frac{\langle {\cal E}\rangle_t }{\langle {\cal E}\rangle_{-\tau}} = 
 \frac{\omega(t)}{\omega_0} (\beta + \Delta\beta), 
\label{E-u-}
\ee
where
\be
 \beta = |u_{+}|^2 + |u_{-}|^2 = 1 + 2|u_{-}|^2,
\label{defbeta}
\ee
\be
\Delta\beta = \left\{ \left[\omega_0^2 \langle x^2 \rangle_{-\tau} - \langle p^2 \rangle_{-\tau}
\right]  \mbox{Re}\left(u_{+} u_{-}\right) +
\omega_0 \langle xp + px \rangle_{-\tau} \mbox{Im}\left(u_{+} u_{-}\right) \right\}
/{\langle {\cal E}\rangle_{-\tau}}.
\label{deltabeta}
\ee

Equation (\ref{E-u-}) can be interpreted as a generalized adiabatic formula for the energy 
after the frequency passes {\em slowly\/} through zero value. 
It shows that the quantum mechanical mean energy
is proportional to the instant frequency $\omega(t)$ in the adiabatic regime. 
However, the proportionality coefficient
strongly depends on the initial conditions in the most general case. This is in agreement
with the classical results shown in Figures \ref{fig-T1-G1000} and \ref{fig-T1-G1000-th}.
For this reason, we concentrate hereafter on 
the important special case when 
\be
\langle p^2\rangle_{-\tau} = \omega_0^2 \langle x^2\rangle_{-\tau}, \qquad 
\langle xp + px \rangle_{-\tau}=0.
\label{special}
\ee
It includes the vacuum, thermal and Fock initial quantum states.
Then, $\Delta\beta =0$. In addition, many formulas can be simplified:
\be
\langle x^2\rangle_t = \omega_0 \langle x^2\rangle_{-\tau} |\vep(t)|^2, \qquad
\langle p^2 \rangle_t  = \langle p^2\rangle_{-\tau} |\dot\vep(t)|^2/\omega_0,
\ee
\be
\langle {\cal E}\rangle_t = \frac{\langle {\cal E}\rangle_{-\tau}}{2\omega_0} 
\left[ \omega^2(t) |\vep(t)|^2 + |\dot\vep(t)|^2\right].
\label{Etvac}
\ee

In principle, the choice of the initial point of integration  in Equation (\ref{t*}), defining the phase function $\phi(t)$
in Equation (\ref{adsol+}),
can be arbitrary, since it influences the phases of coefficients $u_{\pm}$ only. However, the point $t_*=0$ is distinguished
in our problem,
because $\omega(t_*)=0$. Therefore, we assume hereafter that $t_*=0$ in the definition of the
phase $\phi(t)$ (\ref{t*}).

Note that after the Lewis and Riesenfeld paper \cite{LR}, many authors working on various problems related to the harmonic
oscillator with a time-dependent frequency used as a starting point not the linear equation (\ref{eq}) but its nonlinear
analog (known under the name ``Ermakov equation'')
\be
\ddot\rho + \omega^2(t)\rho = \rho^{-3}, \qquad \rho(t) \equiv |\vep(t)|,
\label{Erm}
\ee
which follows from (\ref{eq}) if one writes $\vep=\rho \exp(i\chi)$ and takes into account the condition (\ref{Wr}).
Then, one can rewrite Equation (\ref{Etvac}) as follows,
\be
\langle {\cal E}\rangle_t = \frac{\langle {\cal E}\rangle_{-\tau}}{2\omega_0} 
\left\{ \omega^2(t) \rho^2(t) + [\dot\rho(t)]^2 + [\rho(t)]^{-2} \right\} \ge 
\frac{\langle {\cal E}\rangle_{-\tau}}{2\omega_0} 
\left\{ 2|\omega(t)|  + [\dot\rho(t)]^2  \right\}.
\label{Etvacrho}
\ee
Consequently, the mean energy always {\em increases\/} when the frequency returns to its initial value,
unless the time derivative $\dot\rho(t)$ is negligibly small, i.e., if $u_{-} \neq 0$  [under the conditions (\ref{special})].
Many references on the subjects related to the Ermakov equation can be found, e.g., in the review \cite{Sus10} and a recent paper
\cite{Ramos18}.
However, we prefer to use the linear equation (\ref{eq}), because the keys that help us to solve the adiabatic problem are the coefficients
$u_{\pm}$ in the asymptotic formula (\ref{adsol+}).

But how to find these constant coefficients $u_{\pm}$? Numeric results of Section \ref{sec-clas}
(especially Figure \ref{fig-meanE})
indicate that the answer depends on the exponent $n$ in the form of the frequency transition
through zero: $\omega^2(t) \sim |t|^n$ when $t\to 0$ (assuming that $\omega(t)=0$ at $t=0$).
It is remarkable that the explicit dependence of $|u_{-}|^2$ on the index $n$
can be found analytically, as shown in the next section.

\section{Exact solutions for the power profile of the frequency}
\label{sec-power}

It is known  that Equation (\ref{eq}) with the time-dependent frequency 
$\omega^2(t) =\omega_0^2 |t/\tau|^n$
can be reduced to the Bessel equation 
\be
\frac{d^2 Z}{dy^2} + \frac{1}{y}\frac{d Z}{dy} +\left(1- \frac{\nu^2}{y^2}\right) Z =0
\label{eqBes}
\ee
for $t>0$ (see, e.g., papers \cite{Lew68,Kim94}). The same can be done for $t<0$,
as soon as the initial equation is invariant with respect to the time reflection $t\to -t$.
One can verify that Equation (\ref{eq}) goes to (\ref{eqBes}) 
 with the aid of the following transformations:
\be
x(t) = \sqrt{|t|}\,Z[y(t)], \qquad \nu = \frac{1}{n+2}, \quad y(t) = g \left|\frac{t}{\tau}\right|^{\gamma}, \quad
\gamma = \frac{1}{2\nu}, \quad g=2G\nu, \quad G=\omega_0\tau.
\label{transZ}
\ee
Hence, the function $\vep(t)$ can be written as a superposition of the Bessel functions $J_{\nu}(y)$ and $J_{-\nu}(y)$,
although with different coefficients in the regions of $t<0$ and $t>0$:
\be
\vep(t) = \sqrt{|t|}\times \left\{
\begin{array}{ll}
\left\{A_{-} J_{\nu}[y(t)] + B_{-} J_{-\nu}[y(t)]\right\}, & t<0
\\
\left\{A_{+} J_{\nu}[y(t)] + B_{+} J_{-\nu}[y(t)]\right\}, & t>0
\end{array} \right. .
\label{vepBes}
\ee
Constant complex coefficients $A_{-}$ and $B_{-}$ can be found from the initial conditions (\ref{invep}).
Remembering that $d|t|/dt =-1$ for $t<0$, one obtains the following equations:
\[
A_{-} J_{\nu}\left(g \right) + B_{-} J_{-\nu}\left(g \right) = 1/\sqrt{G},
\]
\[
A_{-} J^{\prime}_{\nu}\left(g \right) + B_{-} J^{\prime}_{-\nu}\left(g \right) 
=  -\,\frac{1}{\sqrt{G}} \left( i + \frac{1}{2G}\right),
\]
where $J^{\prime}_{\pm\nu}(z)$ means the derivative of the Bessel function $J_{\pm\nu}(z)$ with respect to its argument $z$.
Using the known Wronskian \cite{Grad,BE}
\[
J_{\nu}(z)J^{\prime}_{-\nu}(z) - J_{-\nu}(z)J^{\prime}_{\nu}(z) = -2\sin(\nu\pi)/(z\pi),
\]
we obtain the following expressions:
\[
A_{-} = -\,\frac{\nu \pi \sqrt{G}}{\sin(\nu\pi)}\left[ J^{\prime}_{-\nu}\left(g \right)
+ \left( i + \frac{1}{2G}\right)J_{-\nu}\left(g \right) \right],
\]
\[
B_{-} = \frac{\nu \pi \sqrt{G}}{\sin(\nu\pi)}\left[ 
 \left( i + \frac{1}{2G}\right)J_{\nu}\left(g \right)
 + J^{\prime}_{\nu}\left(g \right) \right].
\]
Using the known identities (see, e.g., formulas 7.2 (54) and 7.2 (55) in \cite{BE})
\be
J_{\nu}(z) \pm \frac{z}{\nu}J^{\prime}_{\nu}(z) = \frac{z}{\nu}J_{\nu \mp 1}(z),
\label{idenJnu}
\ee
we can simplify formulas for the coefficients $A_{-}$ and $B_{-}$:
\be
A_{-} = \frac{\nu \pi \sqrt{G}}{\sin(\nu\pi)}\left[ 
J_{1-\nu}\left(g \right) -i J_{-\nu}\left(g \right) 
\right],
\qquad
%\be
B_{-} = \frac{\nu \pi \sqrt{G}}{\sin(\nu\pi)}\left[ 
 i J_{\nu}\left(g \right)
 +J_{\nu -1}\left(g \right) \right].
 \label{A-B-}
 \ee
The time derivative of function (\ref{vepBes}) at $t<0$ (when $d|t|/dt = -1$) can be written with the aid of identities  (\ref{idenJnu}) as follows:
\be
d\vep/dt = \frac{y}{2\nu\sqrt{|t|}}\left[  B_{-} J_{1-\nu}(y) - A_{-} J_{\nu-1}(y)\right], \qquad t\le 0.
\ee
On the other hand,
\be
d\vep/dt = \frac{y}{2\nu\sqrt{t}}\left[  A_{+} J_{\nu-1}(y) -  B_{+} J_{1-\nu}(y) \right], \qquad t\ge 0.
\ee

Using the leading term of the Bessel function $J_p(z) = z^p/[2^p \Gamma(p+1)]$ at $z \to 0$,
 one can see that $\sqrt{|t|}J_{\nu}(y) \to 0$ when $t\to 0$,
while the product $\sqrt{|t|}J_{-\nu}(y)$ tends to a finite value in this limit. 
Consequently, the continuity of function $\vep(t)$ 
at $t=0$ implies the condition $B_{+}=B_{-}$. On the other hand, 
$y J_{1-\nu}(y)/\sqrt{|t|} \to 0$ at $t\to 0$, while 
the product $yJ_{\nu-1}(y)/\sqrt{|t|}$ tends to a finite value in this limit. 
Hence, the continuity of derivative $d\vep/dt$ at $t=0$
can be guaranteed under the condition $A_{+} = -A_{-}$.
Then, one can verify that the Wronskian identity (\ref{Wr}) 
is satisfied identically, both for $t\le 0$ and $t\ge 0$, in view of the identity \cite{BE}
\be
J_{\nu}(z) J_{1-\nu}(z) + J_{-\nu}(z) J_{\nu-1}(z) = 2\sin(\nu\pi)/(z\pi).
\ee

\begin{figure}[hbt]
\centering
\includegraphics[scale=0.35]{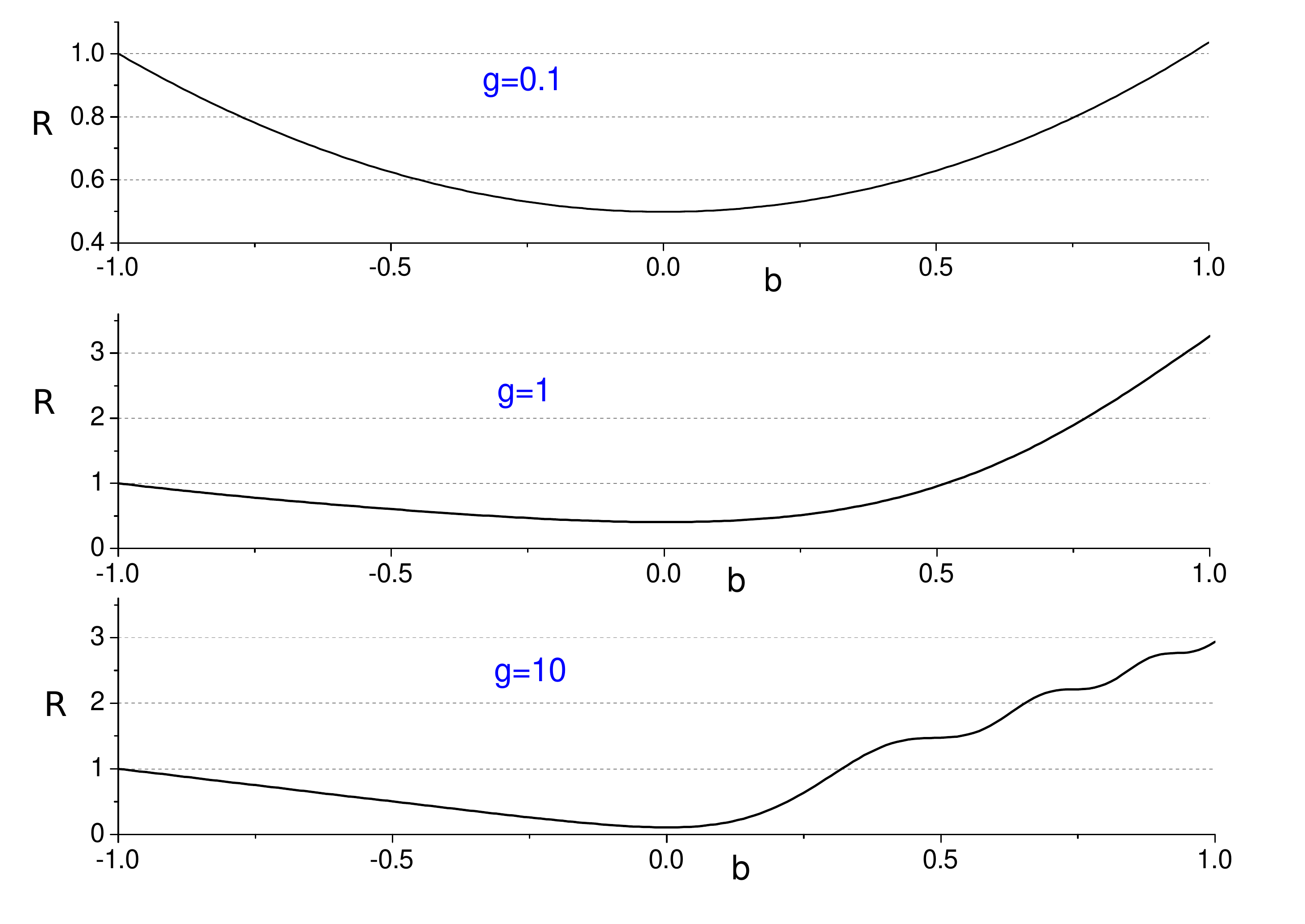}
\caption{The function $R(b)$ for $g = 0.1, 1.0, 10$ and $\nu = 1/4$.
}
\label{fig-Rt}
\end{figure}

Using Equations (\ref{vepBes}) and (\ref{A-B-}), one can write down formula (\ref{Etvac}) for
the mean energy ratio ${\cal R}(t) = {\cal E}(t)/{\cal E}(-\tau)$ 
as follows:
\be
\left. 
\begin{array}{l}
{\cal R}(t<0) 
\\
{\cal R}(t>0)
\end{array}
\right\}
= \frac{1}{8} \left[\frac{g\pi}{\sin(\nu\pi)}\right]^2\left|\frac{t}{\tau}\right|^{n+1}
 \left[K_{-}(g)K_{+}(y) + K_{+}(g)K_{-}(y) \mp 2 K_{0}(g)K_{0}(y)  \right],
\label{RKKK}
 \ee
where
\[
K_{+}(z) =  J_{\nu}^2(z) + J_{\nu-1}^2(z), \quad K_{-}(z) =  J_{-\nu}^2(z) + J_{1-\nu}^2(z), \quad
K_0(z) = J_{\nu-1}(z) J_{1-\nu}(z) - J_{\nu}(z) J_{-\nu}(z).
\]
Figure \ref{fig-Rt} shows the function $R(b)$, where $-1\le b \equiv t/\tau \le 1$, for 
$\nu=1/4$ (i.e., $n=2$ and $y = gb^2$) and three values
of parameter $g = 0.1, 1.0, 10$.

\begin{figure}[hbt]
\centering
\includegraphics[scale=0.35]{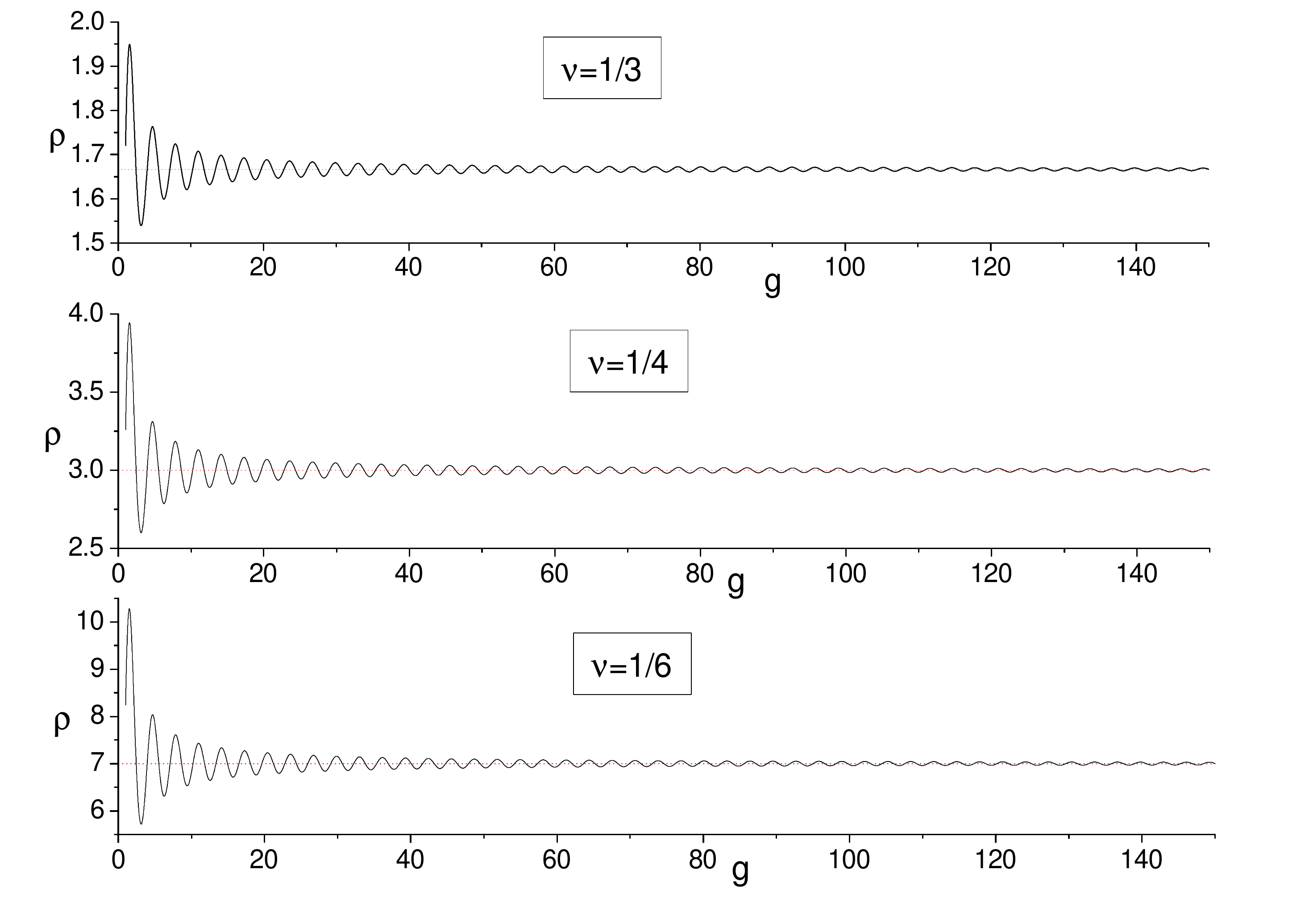}
\caption{The function $\rho(g)$ (\ref{rhog}) for $g >1$ and $\nu = 1/3, 1/4, 1/6$.
}
\label{fig-Rg}
\end{figure}

Using the known formulas
\be
J_{\nu}(y) \approx (y/2)^{\nu}/\Gamma(\nu +1), \qquad y \ll 1,
\label{Jnuy0}
\ee
\be
\Gamma(z)\Gamma(1-z) = \pi/\sin(\pi z),
\label{Gamz1-z}
\ee
one can see that $R(b)$ is totally symmetric in the limit $g\to 0$ (an instantaneous frequency jump through
zero value), when $R(b) = \left(1 + |b|^n\right)/2$.
However, the symmetry is broken for not very small values of parameter $g$.
The known asymptotic formula for the Bessel functions of large arguments \cite{Grad,BE},
\be
J_{\nu}(z) \sim \sqrt{\frac{2}{\pi z}}\cos\left(z - \frac{\nu\pi}{2} - \frac{\pi}{4}\right),
\ee
results in the following simple expressions for $z \gg 1$:
\[
K_{\pm}(z) = \frac{2}{\pi z}, \qquad K_0(z) = \frac{2 \cos(\nu\pi)}{\pi z}.
\]
Hence, in the adiabatic limit ($g\gg 1$ and $y \gg 1$), we obtain 
\[
\left. 
\begin{array}{l}
{\cal R}(t<0) 
\\
{\cal R}(t>0)
\end{array}
\right\}
= \frac{\omega(t) \left[1 \mp \cos^2(\nu\pi)\right]}{\omega_0\sin^2(\nu\pi)}.
 \]
If $t <0$, we arrive exactly at the adiabatic formula (\ref{inv}) for any value of the power $n$. 
On the other hand, if $t>0$
(i.e., after the frequency passed through zero value),  we see again the linear proportionality 
\be
{\cal R}(t) = \beta \omega(t)/\omega_0, \qquad
\beta = \frac{1 + \cos^2(\nu\pi)}{\sin^2(\nu\pi)}, \qquad \nu = \frac{1}{n+2}.
\label{Rnu}
\ee
The proportionality coefficient $\beta$ depends on parameter $n$.
For example, $\beta = 5/3$ for $n=1$, $\beta = 3$ for $n=2$, and $\beta = 7$ for $n=4$, 
in %full 
accordance with Figure \ref{fig-meanE}.
If $n \ll 1$, then $\beta \approx 1$, while $\beta \approx 2(n/\pi)^2$ for $n \gg 1$.
To see the limitations on the validity of the adiabatic approximation $g\gg 1$, we plot
in Figure \ref{fig-Rg} the ratio $\rho \equiv{\cal R}(\tau)$, 
\be
\rho =\frac{1}{4} \left[\frac{g\pi}{\sin(\nu\pi)}\right]^2
 \left[K_{-}(g)K_{+}(g) +  K_{0}^2(g)\right],
\label{rhog}
\ee
as function of $g$ for $\nu = 1/3, 1/4, 1/6$.
We see that the generalized adiabatic approximation (\ref{Rnu}) has the accuracy better than $1$\%
for $g>100$.

In view of formula (\ref{Jnuy0}),
the only nonzero contribution to the right-hand side of Equation (\ref{RKKK}) at $t \to 0$
is given by the function 
$K_{+}(y) \approx J_{\nu-1}^2(y) \approx 
\left[ (y/2)^{\nu-1}/\Gamma(\nu)\right]^2 \sim |t|^{-(n+1)}$,
whereas  the contributions of $K_{-}(y) \sim |t|^{-1}$ and $K_{0}(y) \sim |t|^0$ are eliminated
by the term $|t|^{(n+1)}$. Hence,
\be
{\cal R}(t=0) = \frac{\pi g^{2\nu-1}}{\left[ 2^{\nu} \Gamma(\nu) \sin(\pi\nu)\right]^2}, \qquad g \gg 1.
\ee
This means, in particular, that the adiabatic formula (\ref{Rnu}) holds under the condition
$\omega(t)/\omega_0 \gg g^{2\nu-1}$
(provided $g =2\nu \omega_0 \tau \gg 1$).

\section{Transition rules for adiabatic coefficients after frequency passes through zero value}
\label{sec-trans}

Equation (\ref{Rnu}) means that parameter $|u_{-}|$ [determining the adiabatic evolution of the mean
energy after the frequency passes through zero value according to formula (\ref{E-u-})]
has the following form:
\be
|u_{-}| = \cot(\nu\pi) = \cot\left(\frac{\pi}{n+2}\right).
\label{u-nu}
\ee
This formula can be derived directly from Equation (\ref{vepBes}). If $g \gg 1$, 
Equation (\ref{A-B-}) assumes the following asymptotic form:
\be
A_{-} \approx \frac{\sqrt{\nu\pi}}{\sin(\nu\pi)} 
\exp\left[i\left(g + \frac{\nu\pi}{2} -\frac{3\pi}{4}\right)\right], \qquad
B_{-} \approx \frac{\sqrt{\nu\pi}}{\sin(\nu\pi)} 
\exp\left[i\left(g - \frac{\nu\pi}{2} +\frac{\pi}{4}\right)\right]
\label{A-B-as}
\ee
Then, Equation (\ref{vepBes}) results in the following expressions for $y \gg 1$:
\be
\vep(t<0) \approx [\omega(t)]^{-1/2} e^{i(g-y)}, 
\quad
\vep(t>0) \approx \frac{e^{ig} }{ [\omega(t)]^{1/2} \sin(\nu\pi)}
 \left[ e^{iy} + i\cos(\nu\pi) e^{-iy} \right].
\label{vep-as-0}
\ee
On the other hand, calculating the phase $\phi(t)$ according to the definition (\ref{t*}),
we find  $\phi = -y$ for $t<0$ and ${\phi} = y$
for $t>0$. Hence, omitting the common phase term $e^{ig}$ in Equation (\ref{vep-as-0})
and comparing this equation with (\ref{adsol}) and (\ref{adsol+}),
we obtain the following expressions for the coefficients $u_{\pm}$:
\be
u_{+} = [\sin(\nu\pi)]^{-1}, \qquad u_{-} = i\cot(\nu\pi).
\label{upm-nu}
\ee
They satisfy exactly the identity (\ref{uvcond}) and result in formula (\ref{u-nu}).
Note that coefficient $u_{+}$ given by Equation (\ref{upm-nu}) is real. However, probably,
the reality of this coefficient is due to the specific exact power shape of function
$\omega(t)$ considered in this section.  
For other functions $\omega(t)$ with a similar behavior when $\omega \to 0$, 
this coefficient can be complex, although with the same absolute value.
An example is given in the next section.
However, the formulas for the absolute values $|u_{\pm}|$,
\be
|u_{+}| = [\sin(\nu\pi)]^{-1}, \qquad |u_{-}| = \cot(\nu\pi),
\label{absupm-nu}
\ee
seem to be universal after a single frequency passage through zero.

\section{Exact solution for the tanh profile of the frequency}
\label{sec-EE}

An interesting example of exact solutions corresponds to the time-dependent frequency
(a special case of the family of Epstein--Eckart profiles \cite{Eckart,Epstein})
\be
\omega^2(t) =\omega_0^2 \tanh^2(\kappa t/2),
 \quad -\infty < t < \infty, \quad \kappa >0.
\label{omtanh}
\ee
In this case, solutions to Equation (\ref{eq}) can be written  in terms of the Gauss hypergeometric function 
\[
F(a,b;c;x) = \sum_{n=1}^{\infty} \frac{(a)_n (b)_n x^n}{(c)_n n!},
\]
satisfying the equation
\be
x(1-x)F^{\prime\prime} +(c-(a+b+1)x)F^{\prime} -abF =0.
\label{eqF}
\ee
The first step to come to Equation (\ref{eqF})  is to introduce the new variable
$\xi = \tanh(\kappa t/2)$. Then, Equation (\ref{eq}) takes the form
\be
(1-\xi^2)^2 \frac{d^2 x}{d\xi^2} - 2\xi (1-\xi^2) \frac{d x}{d\xi} + 
4 \tilde{\omega}_0^2 \xi^2 x =0, \qquad \tilde\omega_0 \equiv \omega_0/\kappa.
\ee
We wish to arrive to the function $F(a,b;c;y)$ with $y=(1+\xi)/2$. In such a case,
$y=0$ when $t= -\infty$, so that the initial condition (\ref{invep}) can be easily satisfied,
as soon as $F(a,b;c;0) =1$. On the other hand, there are many relations for the function
 $F(a,b;c;1)$, which arises when $t\to \infty$. Then, the asymptotics of function $\vep(t)$
can be easily found.
Therefore, looking for the solution in the form $x(t) = [y(1-y)]^{^\alpha} f(y)$,
we obtain the equation
\be
y^2(1-y)^2 f^{\prime\prime} + y(1-y)(1-2y)(1+2\alpha) f^{\prime}
+\left[(2y-1)^2\left(\alpha^2 + \tilde{\omega}_0^2\right) - 2\alpha y(1-y)\right] f =0.
\ee
Consequently, choosing $\alpha = i \tilde\omega_0$, we arrive at the solution
\be
\vep(t) = \omega_0^{-1/2}[y(1-y)]^{i \tilde\omega_0} F\left(a_{+}, a_{-}; c; y\right) =
\omega_0^{-1/2}[2\cosh(\kappa t/2)]^{-2i \tilde\omega_0} F\left(a_{+}, a_{-}; c; y\right),
\label{solth}
\ee
where
\be
y =  \frac12[1+ \tanh(\kappa t/2)] = \left(1 + e^{-\kappa t}\right)^{-1},
\ee
\be
a_{\pm} = \frac12 + 2i \tilde\omega_0 \pm r, \qquad r = \frac12\sqrt{1-16 \tilde\omega_0^2}, \qquad
c = 1 + 2i \tilde\omega_0.
\ee

If $t\to -\infty$, function (\ref{solth}) goes to 
$\vep(t) = \omega_0^{-1/2}\exp(i\omega_0 t)$.
If $t\to \infty$ (and $y \to 1$), we can use the analytic continuation of the hypergeometric
function, given, e.g., by formula 2.10(1) from \cite{BE},
\beqn
F(a,b;c;y) &=& \frac{\Gamma(c) \Gamma(c-a-b)}{\Gamma(c-a) \Gamma(c-b)} F(a,b; a+b+1-c; 1-y)
\nonumber \\
&& 
+ \frac{\Gamma(c) \Gamma(a+b-c)}{\Gamma(a) \Gamma(b)} (1-y)^{c-a-b} F(c-a,c-b; c+1-a-b; 1-y).
\eeqn
Then, function (\ref{solth}) assumes the asymptotic form
\be
\vep(t) = \omega_{0}^{-1/2}\left[v_{+} e^{i\omega_{0}t} + v_{-} e^{-i\omega_{0}t}\right],
\label{uvsol}
\ee
\be
v_{+} = \frac{\Gamma(1 + 2i\tilde\omega_0)\Gamma(2i\tilde\omega_0)}
{\Gamma(1/2 + 2i\tilde\omega_0 +r)\Gamma(1/2 + 2i\tilde\omega_0 -r)} , \qquad
v_{-} = \frac{\Gamma(1 + 2i\tilde\omega_0)\Gamma(-2i\tilde\omega_0)}
{\Gamma(1/2  +r)\Gamma(1/2  -r)}.
\ee
Using the relation (\ref{Gamz1-z}),
we can simplify the expression for coefficient $v_{-}$:
\be
v_{-} = 
\frac{i \cos\Big(\pi \sqrt{1/4 - 4\tilde\omega_0^2}\Big)}
{\sinh(2\pi\tilde\omega_0)}.
\label{v--f}
\ee 
The quantity $|v_{-}|^2$ increases with increase of $\tilde\omega_0$. If $\tilde\omega_0= 1/4$, then
$|v_{-}|^2 = [\sinh(\pi/2)]^{-2} \approx 0.19$.
In the adiabatic limit $\tilde\omega_0 \gg 1$  we have
$v_{-} \approx i\coth(2\pi\tilde\omega_0)$, i.e., $v_{-} \to i$ and
 $|v_{-}| \to 1$ when $\tilde\omega_0 \to \infty$.
In this limit, we have 
$r = 2i \tilde\omega_0 +{\cal O}(\tilde\omega_0^{-1})$. Then, we can write
\[
v_{+} \approx \frac{ 2i\tilde\omega_0[\Gamma(2i\tilde\omega_0)]^2}
{\sqrt{\pi}\Gamma(1/2 + 4i\tilde\omega_0 )}.
\]
Using the asymptotic Stirling formula for the Gamma function, 
\be
\Gamma(z) \approx \sqrt{2\pi}\exp\left[(z-1/2)\ln(z) -z\right], \quad |z| \gg 1.
\label{Stirling}
\ee
we obtain the expression
\be
v_{+} \approx \sqrt{2}\, \exp\left[ -4i\tilde\omega_0 \ln(2)\right].
\ee
Consequently, $|v_{+}|^2 - |v_{-}|^2 =1$ and ${\cal E}(\infty)/{\cal E}(-\infty) =3$,
in accordance with formula (\ref{Rnu}).

The asymptotic form (\ref{uvsol}) is similar to the general adiabatic solution (\ref{adsol+}).
Using the definition (\ref{t*}) of the phase $\phi(t)$, we obtain the formula 
[remember that $\omega(t) = |\tanh(\kappa t/2)|$ in this section, as soon as function
$\omega(t)$ is assumed to be non-negative in formula (\ref{adsol+})] 
\be
\phi(t) = 2\tilde\omega_0 \ln[\cosh(\kappa t/2)] \mbox{sign}(t).
\ee
If $t \to \pm \infty$, then, $\phi(t) \approx \omega_0 t -2\tilde\omega_0 \ln(2) \mbox{sign}(t)$.
This means that, according to (\ref{adsol+}), the function
$\omega_0^{-1/2} \exp\left(i\left[\omega_0 t + 2\tilde\omega_0 \ln(2)\right]\right)$
at $t \to -\infty$ goes to the following superposition at $t \to \infty$:
\[
\omega_0^{-1/2} \left\{ u_{+}\exp\left(i\left[\omega_0 t - 2\tilde\omega_0 \ln(2)\right]\right)
+ u_{-}\exp\left(i\left[-\omega_0 t + 2\tilde\omega_0 \ln(2)\right]\right)\right\}.
\]
Comparing this expression with (\ref{uvsol}), we conclude that
\be
u_{+}=v_{+} = \sqrt{2}\, \exp\left[ -4i\tilde\omega_0 \ln(2)\right], \qquad
u_{-}=v_{-} = i.
\ee
We see that the phases of complex coefficients $u_{\pm}$ are sensitive to the rate of the
adiabatic evolution through the term $\tilde\omega_0$. A strong consequence of this 
result is considered in the next section.

\section{Double adiabatic passage of frequency through zero value}
\label{sec-double}

What can happen if the frequency will pass again through zero value? Then, we have
to make the transformation of function (\ref{adsol+}), using the superposition principle
and two additional observations.
First: the function $\omega^{-1/2} \exp(-i\phi)$ transforms as function $\vep^*(t)$ after
the frequency passes through zero value.
Second: applying the transformation rule (\ref{adsol+}) to the second transition, we 
must use the phase $\tilde\phi(t)$, where the integral over frequency is taken from
the second transition point $t_{**}$.
Obviously, 
\be
\phi(t) = \tilde\phi(t) + \Phi, \qquad 
\Phi(t_*, t_{**}) = \int_{t_*}^{t_{**}} \omega(z)dz.
\label{Phi}
\ee
We suppose that the transition rule through the second zero has the form
\[
\left\{[\omega(t)]^{-1/2}e^{i\tilde\phi(t)}\right\}_{t<t_{**}} \to
\left\{[\omega(t)]^{-1/2}\left[w_{+}e^{i\tilde\phi(t)} + w_{-}e^{-i\tilde\phi(t)}\right]
\right\}_{t>t_{**}}, \quad |w_{+}|^2 - |w_{-}|^2 =1.
\]
Comparing two forms of the solution $\vep(t)$ for $t > t_{**}$ (far enough from that point)
we arrive at the equality
(omitting the common term $\omega^{-1/2}$ and using the notation $U_{\pm}$ for the 
coefficients at $t > t_{**}$)
\[
U_{+}e^{i\tilde\phi(t)} + U_{-}e^{-i\tilde\phi(t)} =
u_{+}e^{i\Phi}\left[w_{+}e^{i\tilde\phi(t)} + w_{-}e^{-i\tilde\phi(t)}\right]
+ u_{-}e^{-i\Phi}\left[w_{+}^*e^{-i\tilde\phi(t)} + w_{-}^*e^{i\tilde\phi(t)}\right].
\]
Hence, 
\be
U_{+} = w_{+} u_{+}e^{i\Phi} + w_{-}^* u_{-}e^{-i\Phi}, \qquad
U_{-} = w_{-} u_{+}e^{i\Phi} + w_{+}^* u_{-}e^{-i\Phi}.
\label{Upm}
\ee
One can verify that the identity $|U_{+}|^2 - |U_{-}|^2 =1$ is fulfilled exactly. 
The adiabatic mean energy amplification factor
after the second passage through zero frequency equals
\be
\beta = 1 +2|U_{-}|^2
 = 1 + 2\left[|w_{-}|^2 |u_{+}|^2 + |w_{+}|^2 |u_{-}|^2
+ 2\mbox{Re}\left(w_{+}w_{-}u_{+}u_{-}^* e^{2i\Phi}\right)\right].
\label{beta**}
\ee
%\frac{8\cos^2(\nu\pi)}{\sin^4(\nu\pi)}.
In the adiabatic regime, $\Phi \gg 1$. Moreover, this phase is very sensitive to the form of function
$\omega(t)$ and the distance between the zero-point instances $t_{*}$ and $t_{**}$. In addition,
coefficients $u_{\pm}$ and $w_{\pm}$ can be strongly phase-sensitive, as shown in Section \ref{sec-EE}.
This means that it is practically impossible to predict the energy mean value after 
twice zero frequency crossing
 (quite differently from the single crossing).
The extremal values of $\beta$ are as follows,
\be
\beta_{min} = 1 +2 \left(|w_{+}u_{-}| - |w_{-}u_{+}|\right)^2, \qquad
\beta_{max} = 1 +2 \left(|w_{+}u_{-}| + |w_{-}u_{+}|\right)^2.
\ee
In particular, if $w_{\pm}= u_{\pm}$, then $\beta_{min}=1$, meaning that, in principle, 
the mean energy can return to the initial value after the frequency passes through zero value two times. 
On the other hand, $\beta_{max}=1 +8|u_{+}u_{-}|^2$ under the same conditions. If $n=2$,
$\beta_{max} = 17$.

\subsection{Classical illustrations}

\begin{figure}[hbt]
\centering
\includegraphics[scale=0.5]{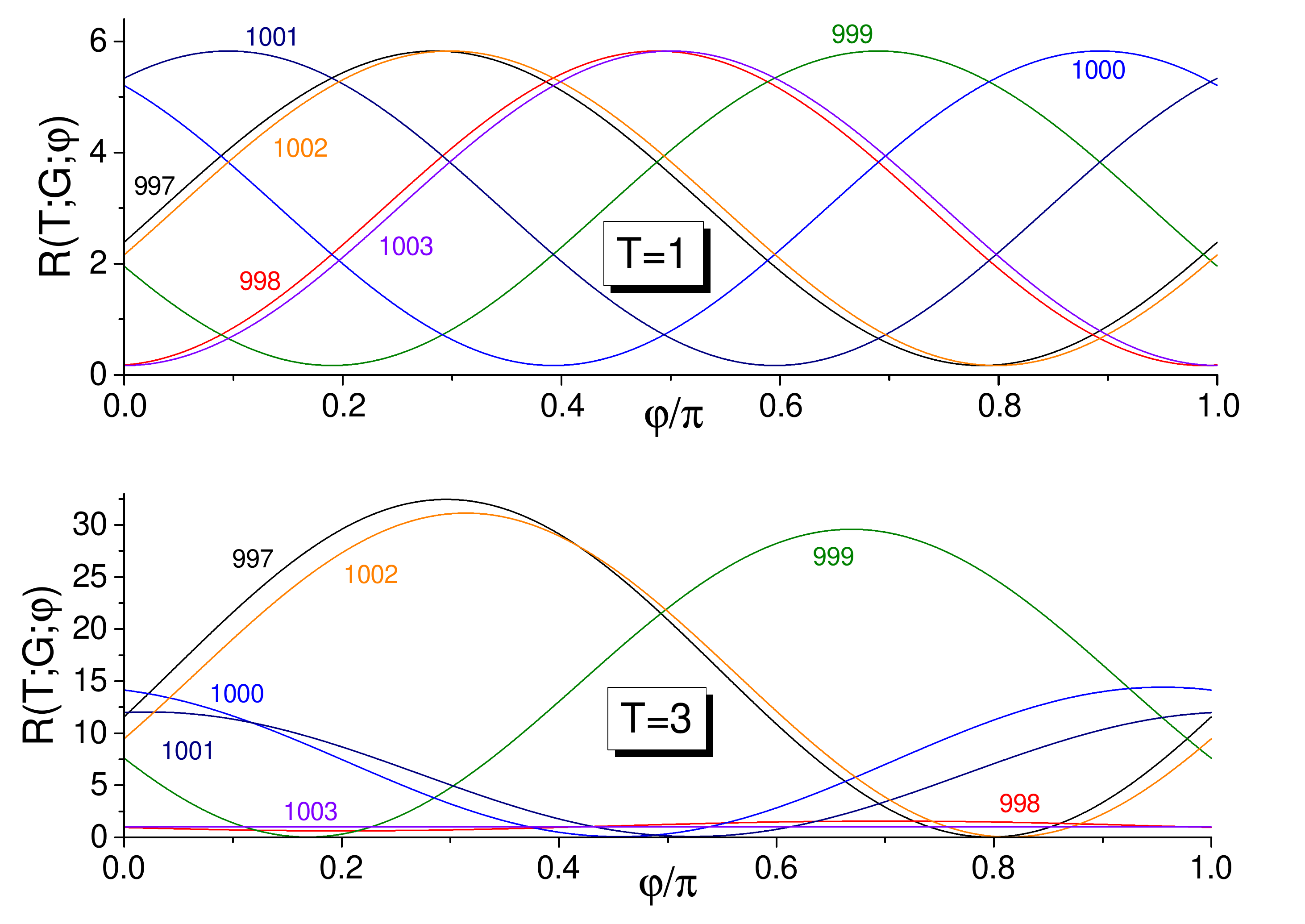}
\caption{The dimensionless energy of a classical particle at the dimensionless instants $T=1$ 
(top) and $T=3$ (bottom) for several different values of the adiabatic paramerer $G =\omega_0\tau$, 
shown near the respective lines.
The frequency profile is $\omega^2(t) = \omega_0^2 \sin^2[\pi t/(2\tau)]|$.
}
\label{fig-sin2}
\end{figure}

To illustrate the effects after the frequency single and double crossings through zero value, 
we considered the classical motion with the frequency
$\omega^2(t) = \omega_0^2 \sin^2[\pi t/(2\tau)]|$ for $t\ge = -\tau$ and the 
initial conditions (\ref{inivf}). Figure \ref{fig-sin2} shows the energy ratio $R$ (\ref{Rf}) 
at the instants $T= t/\tau =1$ and $T=3$, for several values of parameter $G$ close to $G=1000$. 
If $T=1$ (the single crossing), variations of parameter $G$ result in shifts of the curves
without changing the maximal, minimal and average values. On the other hand, the picture
is totally different for $T=3$ (the double crossing).
In this case, $\Phi = 4G/\pi$. Hence, the variation $\Delta \Phi = 2\pi$, when one can expect
a similar behavior, corresponds to $\Delta G = \pi^2/2 \approx 5$. On the other hand, a twice
smaller variation $\Delta G \approx 2$, yielding $\Delta \Phi \approx \pi$, results in the
totally different behavior, as one can see in the figure for $T=3$.

\section{Energy fluctuations}
\label{sec-fluct}

Figures \ref{fig-T1-G1000}, \ref{fig-T1-G1000-th} and \ref{fig-sin2} show strong energy 
fluctuations (as functions of the initial phase) after the frequency passes through zero value.
These fluctuations can be characterized by the variance 
$\sigma_E = \langle E^2\rangle - \langle E\rangle^2$. 
Using the solutions (\ref{solxp}) of the Heisenberg equations of motion, one can write $\sigma_E$
in terms of the fourth- and second-order moments of the canonical variables $x$ and $p$ and 
various products of functions $\vep(t)$, $\dot\vep(t)$ and their complex conjugated partners.
The complete formula is rather cumbersome in the most general case. For this reason,
we consider here the simplest case of the initial Fock quantum state $|N\rangle$. 
Probably, this special case is the most interesting, because the famous adiabatic
theorem in quantum mechanics was proven by Born and Fock \cite{BF} exactly for the Fock states.
In this special case (as well as for arbitrary diagonal mixtures of the Fock states), 
the nonzero statistical moments are those containing {\em even powers\/} of each variable,
$x$ or $p$. After some algebra, one can obtain the following formula (using the dimensionless
variables, assuming $\hbar=m=\omega_0 =1$, so that $\langle x^4 \rangle = \langle p^4\rangle$):
%and $\langle x^2 \rangle = \langle p^2\rangle$):
\be
16\langle E^2\rangle_t = 2\langle x^4 \rangle_{-\tau} \left(A^2 + B^2\right)
+ \langle x^2 p^2 + p^2 x^2\rangle_{-\tau} \left(A^2 - B^2\right)
+ \langle (x p + p x)^2\rangle_{-\tau} C^2,
\ee
where
\[
A(t) = \omega^2(t)|\vep(t)|^2 +|\dot\vep(t)|^2, \quad
B(t) = \mbox{Re}\left[\omega^2(t)\vep^2(t) + {\dot\vep}^2(t) \right], \quad
C(t) = \mbox{Im}\left[\omega^2(t)\vep^2(t) + {\dot\vep}^2(t) \right].
\]
In the adiabatic regime (\ref{adsol+}) we have
\be
A = 2\omega(t)\left(u_{+}|^2 +|u_{-}|^2\right), \quad
B = 4\omega(t) \mbox{Re}\left(u_{+}u_{-}\right), \quad
C = 4\omega(t) \mbox{Im}\left(u_{+}u_{-}\right).
\ee
For the initial Fock state $|N\rangle$ we have
\[
\langle x^4 \rangle_{-\tau} = \frac34 \left( 2N^2 + 2N +1 \right), \quad
\langle x^2 p^2 + p^2 x^2\rangle_{-\tau} = \frac12 \left( 2N^2 + 2N -1 \right), 
\]
\[
\langle (x p + p x)^2\rangle_{-\tau} = 2 \left( N^2 + N +1 \right).
\]
Hence, 
\[
\langle E^2\rangle_t/\omega^2(t) = \left(u_{+}|^2 +|u_{-}|^2\right)^2 (N+1/2)^2
+ 2|u_{+}u_{-}|^2 \left( N^2 + N +1 \right).
\]
Remembering that the mean energy equals 
$\langle E\rangle_t = \omega(t) \left(u_{+}|^2 +|u_{-}|^2\right) (N+1/2)$,
we arrive at the unexpectedly simple formula for the energy variance:
\be
\sigma_E(t) = 2 \omega^2(t) |u_{+}u_{-}|^2 \left( N^2 + N +1 \right), \quad
\frac{\sigma_E(t)}{\langle E\rangle^2_t} = 2 |u_{+}u_{-}|^2 \frac{N^2 + N +1}{N^2 + N +1/4}.
\ee
In the absence of zero frequency values we have $u_{-} =0$. In this case, $\sigma_E(t) \equiv 0$,
in accordance with the Born--Fock theorem. However, this theorem is broken when the frequency
passes through zero value. For example,
for  the initial vacuum state ($N=0$) and the power index $n=2$ of the single frequency
transition through zero value, we obtain 
${\sigma_E(t)}/{\langle E\rangle^2_t} =16$. This ratio can be four times smaller if $N\gg 1$.

In quantum optics, fluctuations are frequently characterized by the {\em Mandel factor\/} \cite{Man-Q}
\be
{\cal Q}= \left(\langle \hat{n}^2 \rangle - \langle \hat{n} \rangle^2\right)/\langle \hat{n} \rangle -1,
 \qquad \hat{n} \equiv \hat{E}/(\hbar\omega) -1/2.
\label{def-Mand}
\ee
Then, for the initial Fock state $|N\rangle$ (having $Q=-1$, which means the so called sub-Poissonian statistics),
we obtain the following instantaneous values: 
\[
\hat{n} = N + 2|u_-|^2(N+1/2), \qquad
\langle \hat{n}^2 \rangle - \langle \hat{n} \rangle^2 = \sigma_E /(\hbar\omega)^2.
\]
Consequently,
\be
Q = \frac{|u_-|^2 \left(1 + 2|u_-|^2\right) +N \left(2|u_-|^4 -1\right) + 2N^2 |u_{+}u_-|^2}
{N + 2|u_-|^2(N+1/2)}.
\label{Q-u-}
\ee
In particular, for $|u_-|^2 =1$ we have
\be
Q = \frac{3 + N + 4N^2}{3N +1}.
\ee
This means that the statistics become super-Poissonian after the frequency passage through zero value. 
However, the super-Poissonianity is not very strong, because $Q \approx (4/3)N \approx (4/9)\langle \hat{n} \rangle$ for $N \gg 1$,
whereas $Q = \langle \hat{n} \rangle$ for the ``strongly super-Poissonian'' thermal states.

\section{Evolution of the Fock states}   
\label{sec-Fock}

What happens with the initial Fock state $|N\rangle$ when the frequency passes through zero value?
Obviously, it cannot survive, as soon as the mean energy and especially the energy
variance increase substantially. This means that the initial Fock state becomes a superposition
of many Fock states. But what is the width of the new distribution? Is it concentrated
near some distinguished states, or it is very wide and almost uniform, especially when $N\gg 1$?
To answer these questions, one can use general results concerning the quantum harmonic
oscillator with time-dependent frequency \cite{Husimi,PP,LR,MMT70} (the details can be
found, e.g., in the review \cite{183-2}).
Remember that the Fock states $|N\rangle$ are eigenstates of the operator 
$\hat{a}^{\dagger} \hat{a}$, 
where $\hat{a}$ and $ \hat{a}^{\dagger}$ are standard annihilation and creation operators.
When frequency varies with time, operators $\hat{a}$ and $ \hat{a}^{\dagger}$ become
new operators, $\hat{A}$ and $ \hat{A}^{\dagger}$, which are quantum integrals of motion.
The time dependent state $|N\rangle_t$ remains the eigenstate of operator
$\hat{A}^{\dagger} \hat{A}$. As soon as operators $\hat{A}$ and $ \hat{A}^{\dagger}$
maintain their {\em linear\/} form with respect to operators $\hat{x}$ and $ \hat{p}$,
the wave function of state $|N\rangle_t$ maintains its functional form as the product
of some Gaussian exponential by the Hermite polynomial. The explicit form, found in 
\cite{PP,MMT70} (see also \cite{Kim94}), is as follows (in dimensionless units with $\hbar=m =1$),
\be
\langle x|N\rangle_t = \left( N!\,\vep \sqrt{\pi}\right)^{-1/2}
\left(\frac{\vep^*}{2\vep}\right)^{N/2}
\exp\left( \frac{i\dot\vep}{2\vep} x^2\right) H_N\left(\frac{x}{|\vep|}\right),
\label{Nxt}
\ee
where $\vep(t)$ is the solution to Equation (\ref{eq}) satisfying conditions (\ref{invep})
and (\ref{Wr}).
Transition probabilities $|\langle M|N\rangle_t|^2$ between the instantaneous
 Fock state $|M\rangle$
[when $\vep(t)= \omega^{-1/2} \exp(i\omega t)$]
and exact time dependent state $|N\rangle_t$ were calculated in different forms in papers 
\cite{Husimi,PP,LR,MMT70}. In the generalized adiabatic regime (\ref{adsol+}), the results
of \cite{PP,MMT70} can be written in the form (symmetric with respect to $M$ and $N$)
\be
|\langle M|N\rangle_t|^2 = \frac{N_<!}{N_>!} |u_{+}|^{-1}
\left[P_{(M+N)/2}^{|M-N|/2}\left(|u_{+}|^{-1}\right)\right]^2,
\label{prob}
\ee
where $P_j^k(z)$ is the associated Legendre polynomial, $N_<= \mbox{min}(M,N)$,
$N_>= \mbox{max}(M,N)$. Formula (\ref{prob}) holds provided $|M-N|/2$ is an integer; otherwise
the probability equals zero.
Note that the probabilities do not depend on time, as soon as the adiabatic solution (\ref{adsol+})
is valid.

In some cases, it can be convenient to use the expression of the associated Legendre polynomials 
in terms of the Gauss hypergeometric function,
\be
P_n^m(x) = \frac{(-1)^m (n+m)!}{2^m (n-m)! m!} (1-x^2)^{m/2} 
F\left(m-n, m+n+1; m+1; \frac{1-x}{2} \right).
\ee
Then,
\be
|\langle M|N\rangle_t|^2 = \frac{2(N_>)! |u_{-}|^{|M-N|}}{(N_<)!  
\left[\left(\frac{|M-N|}{2}\right)! \right]^2 |2u_{+}|^{|M-N|+1}}
\left[F\left(-N_<\,, N_> +1; \frac{|M-N|}{2} +1; \frac{|u_{+}| -1}{2|u_{+}|} \right)\right]^2.
 \label{probF} 
\ee
In the case of a single frequency crossing through zero value, this formula can be also written as
\be
|\langle M|N\rangle_t|^2 = \frac{(N_>)! \sin(\nu\pi) [\cos(\nu\pi)]^{|M-N|}}{(N_<)!  
\left[\left(\frac{|M-N|}{2}\right)! \right]^2 2^{|M-N|}}
\left[F\left(-N_<\,, N_> +1; \frac{|M-N|}{2} +1; \sin^2[(1-2\nu)\pi/4] \right)\right]^2.
 \label{probFnu}
\ee
Among different special cases, we bring here two formulas:
\be
|\langle 2K|0\rangle_t|^2 = \frac{(2K-1)!! |u_{-}|^{2K}}{ (2K)!! |u_{+}|^{2K+1}}
= \frac{(2K-1)!! }{ (2K)!!} \sin(\nu\pi) [\cos(\nu\pi)]^{2K},
\label{prob02K}
\ee
\be
|\langle N|N\rangle_t|^2 =  |u_{+}|^{-1} \left[P_{N}\left(|u_{+}|^{-1}\right)\right]^2 
= \sin(\nu\pi) \left[P_{N}\left(\sin(\nu\pi)\right)\right]^2,
\label{probNN}
\ee
where $P_N(z) \equiv P_N^0 (z)$ is the usual Legendre polynomial. The first equalities in 
Equations (\ref{prob02K}) and (\ref{probNN}) hold in the most general adiabatic case 
(including multiple frequency crossings through zero value), whereas the second equalities 
are valid for the single crossing.
The distribution (\ref{prob02K}) (describing the evolution of the initial ground state)
 decreases monotonously as function of parameter $K$. 
However, the situation is totally different for other initial Fock states, especially when $N\gg 1$. 

\begin{figure}[hbt]
\centering
\includegraphics[scale=0.35]{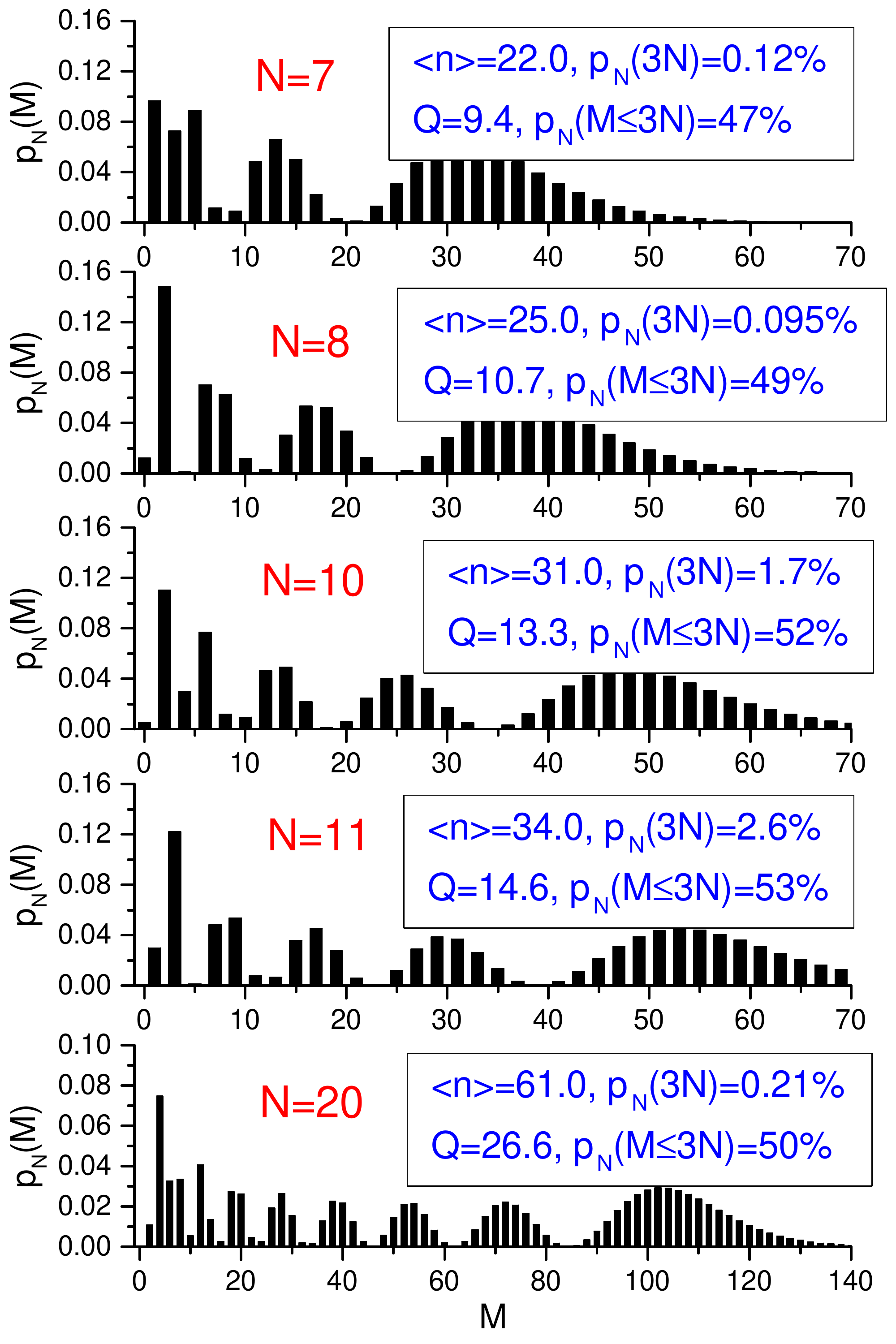}
\caption{The probability $p(M)$ [given by Equation (\ref{probF})]
 of finding the initial Fock state $|N\rangle$ 
in the Fock state $|M\rangle$ 
after the frequency slowly passes through zero value, in the case of 
$|u_{-}|=1$ and $|u_{+}|=\sqrt{2}$.
}
\label{fig-pMN}
\end{figure}

The survival probabilities $p_s(N) \equiv |\langle N|N\rangle_t|^2 $
rapidly diminish with the quantum number $N$. For example,
if $n=2$ ($\nu=1/4$), we find the following surviving probabilities after the single 
frequency crossing zero (when the mean energy triplicates):
\[
p_s(0) = \frac1{\sqrt{2}}, \quad p_s(1) = \frac1{2\sqrt{2}}, \quad 
p_s(2) = \frac1{16\sqrt{2}}, \quad p_s(3) = \frac1{32\sqrt{2}}, \ldots
\]
This means that the initial Fock state $|N\rangle$ 
 becomes a superposition of a large number of different
Fock states $|M\rangle$: see Figure \ref{fig-pMN}.
It is impressive that the probabilty of transition $N\to 3N$ is very small, whereas the probability
$p_N(M\ge 3N)$ is about $50$\%.
In addition, the distribution of probabilities $p_N(M)$ with $M < 3N$ looks rather irregular, whereas
some regular picture is observed for $M>3N$. Unfortunately, we did not succeed to find an analytic approximation
for this regular picture.

\section{Squeezing evolution} 
\label{sec-sqz}  

If the frequency $\omega$ does not depend on time, the evolution of the coordinate variance is given
by the formula
\be
\sigma_x(t) = \sigma_x(0) \cos^2(\omega t) + \frac{\sigma_p(0)}{(m\omega)^2} \sin^2(\omega t)
+ \frac{\sigma_{xp}(0)}{(m\omega)} \sin(2\omega t).
\ee
Minimizing this expression over time, one can write the  minimal value $\sigma_{min}$ as
(similar formulas were obtained, e.g., in papers \cite{LPP88,LPH88,DMP94})
\[
\sigma_{min} = (m\omega^2)^{-1} \left[ E - \sqrt{E^2 -\omega^2 D}\right],
\]
where
\[
E = \sigma_p/(2m) + m\omega^2 \sigma_x/2, \qquad D= \sigma_x \sigma_p - \sigma_{xp}^2.
\]
The quantity $D$ is the simplest example of {\em quantum universal invariants\/} \cite{D-univ},
which do not depend on time (although depend on the initial state) for arbitrary quadratic
Hamiltonians. On the other hand, $D \ge \hbar^2/4$ for any (normalizable) quantum state due to
the Schr\"odinger--Robertson uncertainty relation. The energy of quantum fluctuations $E$
satisfies the inequality $E \ge \hbar\omega/2$. Therefore, it is convenient to use
two dimensionless parameters, $\lambda \ge 1$ and $\gamma\ge 1$, according to the relations
$E =\lambda \hbar\omega/2$ and $D = \gamma^2 \hbar^2/4$.
Then, normalizing the minimal value $\sigma_{min}$
 by the variance in the vacuum %(or coherent)
state $\sigma_{vac} = \hbar/(2m\omega)$, one can obtain the following formula for the
{\em invariant squeezing coefficient\/} $s= \sigma_{min}/\sigma_{vac}$:
\be
s = \lambda - \sqrt{\lambda^2 - \gamma^2} = \frac{\gamma^2}{\lambda + \sqrt{\lambda^2 - \gamma^2}}.
\label{s}
\ee

For the states satisfying the initial conditions (\ref{special}), we have $\gamma =\lambda$
and $s=\lambda \ge 1$. Also, parameter $\lambda$ maintains it initial value 
in the standard adiabatic case (\ref{adiinv}). However, if the frequency passes through
zero value, in the new adiabatic regime (\ref{E-u-}), $\lambda$ goes to $\beta\lambda$,
while $\gamma$ maintains its initial value. Hence, the new squeezing coefficient equals
\be
s= \frac{\lambda}{\beta + \sqrt{\beta^2 -1}}.
\label{snew}
\ee
Hence, the initial vacuum state becomes squeezed when the frequency passes adiabatically 
through zero value. 
Using Equation (\ref{Rnu}), we obtain the following value of the squeezing
coefficient after the single passage through zero:
\be
s = \lambda\tan^2(\nu\pi/2).
\ee
In particular, $s\approx 0.17\lambda$ for $\nu=1/4$ (i.e., $n=2$ and $\beta=3$), 
so that the Fock states $|N\rangle$ become squeezed for $N \le 2$ 
(when $\lambda \le 5$) in this special case.

\section{Conclusions}
\label{sec-concl}

The first main result of the paper is the discovery of the existence of the generalized 
adiabatic invariant in the form of Equation (\ref{E-u-}). In the most general case, the 
adiabatic proportionality coefficient in this equation depends on the initial state. 
This dependence disappears after averaging over 
parameters of families of initial states with the same energy (in particular, such averaging
happens automatically for the initial vacuum, Fock and thermal states). 
Then, universal relations
(\ref{absupm-nu})  exist, {\em provided the frequency passes through zero only once}. 
In the cases of multiple frequency passages through zero, the energy adiabatic coefficients 
become sensitive to the additional parameter - the phase $\Phi$, according to 
Equation (\ref{beta**}). 
As a consequence, the adiabatic behavior after
many crossings through zero frequency value can be quasi-chaotic.
Under specific conditions, the mean energy can return to the
initial value after double frequency passage through zero.
The original Born--Fock adiabatic theorem is broken after the frequency passes through 
zero value. Although the functional shape of the wave function of the initial Fock state
is preserved in the form of the product of a Gaussian exponential by the Hermite polynomial,
the arguments of this form are not determined totally by the instantaneous frequency.
However, the probability distribution over the instantaneous Fock states, determined by the
adiabatic coefficients $|u_{\pm}|$, according to Equation  (\ref{prob}), does not depend
on time, as soon as the adiabatic regime is justified. This statement can be considered as
the generalized Born--Fock theorem; it is the second main result of the paper. 
Note that the time-independent probability distributions
can be different after each frequency passage through zero.
 
In view of the mean energy amplification (e.g., triplication in the most natural case of 
linear frequency dependence near zero point), any initial state becomes significantly
deformed. For example, coherent states (which possess the same quadrature variances as the
vacuum state) will be transformed into squeezed states.
The same can be said about initial thermal states: they will become Gaussian mixed states
with unequal quadrature variances (and squeezed under certain conditions), maintaining the
initial value of the quantum purity.

{
The authors acknowledge the partial support of the Brazilian funding agency 
Conselho Nacional de Desenvolvimento Cient\'{\i}fico e Tecnol\'ogico (CNPq). }

\end{document}